\documentclass[10pt]{article}
\usepackage{multicol}
\usepackage{simpleConference}
\usepackage{arxiv}
\usepackage[backend=bibtex, style=chem-rsc, autocite=superscript]{biblatex}
\usepackage[version=3]{mhchem}
\usepackage{bm} 
\usepackage{float}
\usepackage{fancyhdr} 
\usepackage{comment}

\usepackage[utf8]{inputenc} 
\usepackage[T1]{fontenc}    
\usepackage{hyperref}       
\usepackage{url}            
\usepackage{booktabs}       
\usepackage{amsfonts}       
\usepackage{nicefrac}       
\usepackage{microtype}      
\usepackage{lipsum}		
\usepackage{graphicx}
\usepackage{float}
\usepackage{subfig}
\usepackage{caption}
\usepackage{xcolor}

\newcommand\blfootnote[1]{%
  \begingroup
  \renewcommand\thefootnote{}\footnotetext{#1}%
  \addtocounter{footnote}{-1}%
  \endgroup
}
\usepackage{fontawesome5}
\usepackage{ulem}
\def\dout{\bgroup
 \markoverwith{\lower-0.4ex\hbox
 {\kern-.03em\vbox{\hrule width.2em\kern0.25ex\hrule}\kern-.03em}}%
 \ULon}
\MakeRobust\dout
\usepackage[perpage,hang,bottom]{footmisc}
\makeatletter
\newcommand*{\shifttext}[2]{%
  \settowidth{\@tempdima}{#2}%
  \makebox[\@tempdima]{\hspace*{#1}#2}%
}
\makeatother
\newcommand{\darr}{\shifttext{0.1ex}{$\downarrow$}}
\renewcommand{\thefootnote}{\ifcase\value{footnote}\or \darr  \or \sout{\darr} \or \dout{\darr} \fi}
\usepackage{perpage} 
\MakePerPage{footnote} 

\begin{document}

\title{Modified noise kernels in Gaussian process modelling of energy surfaces}

\author{Fabio E. A.~Albertani\textsuperscript{$\ast \dagger$}\ , \ Alex J. W.~Thom\textsuperscript{$\ast$}}
\thispagestyle{empty}

  	\maketitle
	\begin{abstract}
We explore the use of non homogenous noise kernels  in Gaussian process modelling to improve the potential energy curve models describing stochastic electronic structure data.  We use the same noise kernels on energy curves describing deterministic electronic structure data by creating non-homogenous noise model.  We observe,  as well as agreement between the noise of the model and the stochastic data,  a strong regularisation of the curves when using artificial noises on deterministic data which improves Gaussian processes in the over fitting regime.
	\end{abstract}
	\blfootnote{\textsuperscript{$\ast$}\ Yusuf Hamied Department of Chemistry,  University of Cambridge,  Cambridge,  Lensfield Road,  CB2 1EW}
	\blfootnote{\textsuperscript{$\dagger$}\ fa381@cam.ac.uk}
	\setcounter{footnote}{0}
\begin{multicols}{2}

\section{Introduction}
Gaussian processes (GP),  a popular regression machine learning (ML) framework,  have been extensively used to infer \textit{ab initio} electronic structure data and produce models that allow faster evaluation of computationally expensive data\autocite{Deringer2021} .  One particular type of electronic structure method,  that have not been often used in machine learning,  are stochastic methods. 
\par
As opposed to \textit{deterministic} data,  stochastic data is inherently noisy.  This poses two issues with respect to the learning process: a GP is conventionally trained with a standard white noise,  which assumes homoscedasticity of the training data which is not true if the data is produced by stochastic processes.  Moreover,  the uncertainty distribution of predictions does not take into account the known deviation of the stochastic data. 
\par
\begin{figure}[H]
	\centering
	\includegraphics[width=0.35\textwidth]{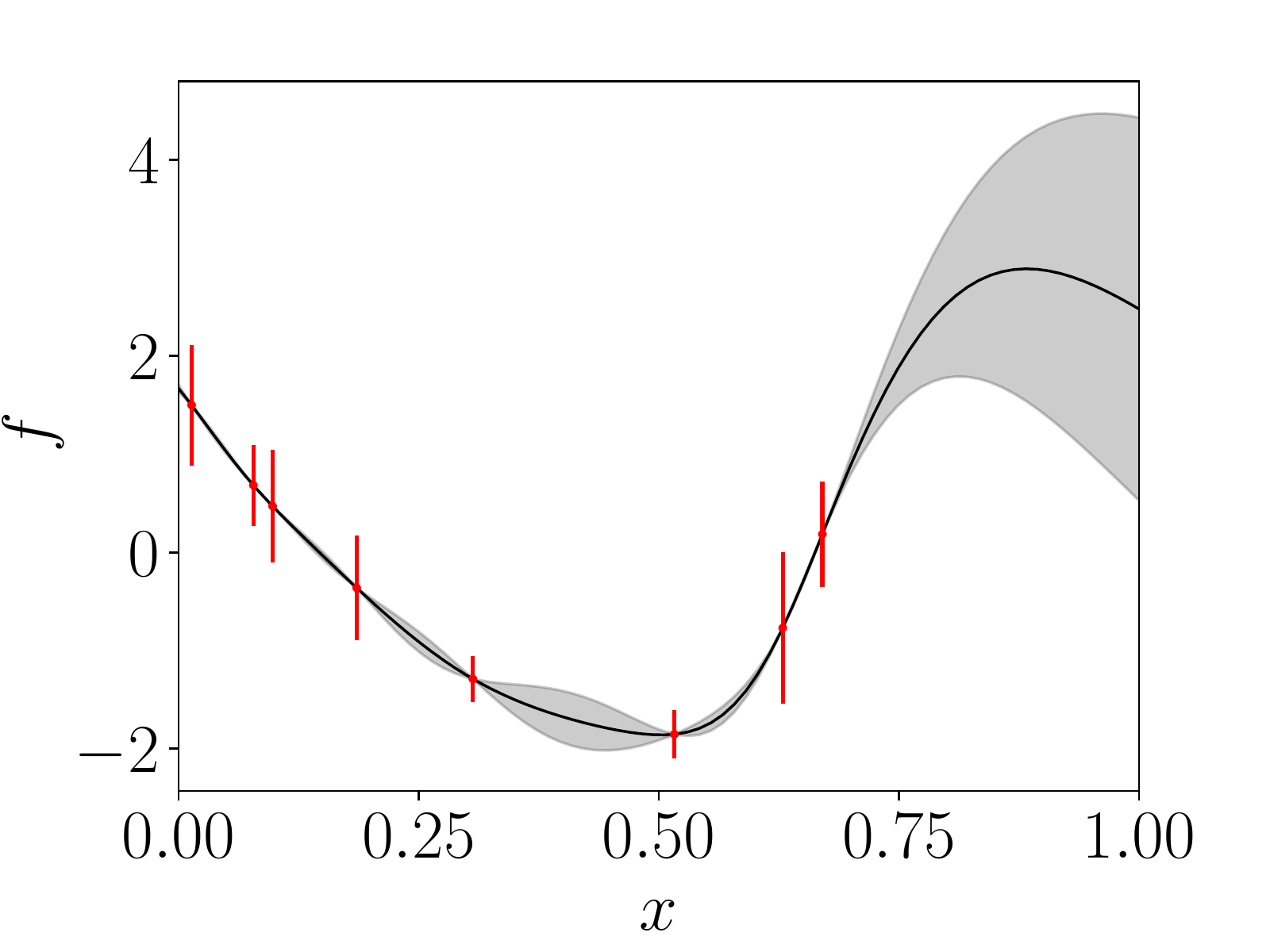}
	\caption{Standard deviation of the latent function of a Gaussian process trained on a dataset.  Assuming that a substantial uncertainty existed on the data,  as shown by the red error bars,  there is a clear discrepancy between the possible real certainty on the target function and the actual machine learning generated certainty.}
	\label{fig:problem}
\end{figure}

As an introductory example,  a set of Morse-potential like data points with randomly generated heteroscedastic noise,  shown in figure \ref{fig:problem}, are used to illustrate the discrepancy that can exist between the two uncertainties.  Moreover,  despite the ML model function showing a seemingly good fit in this case,  noisy data is usually harder to fit as it roughens the model predicted function. 
\par
We explore the use of weighted white kernels,  which have been used to model trust in data of various sources\autocite{LI2020162,Venanzi_Rogers_Jennings_2013,10.2307/26542520,Patel2022},  to improve the modelling of heteroscedastic data and showcase the results on some simple datasets derived from stochastic electronic structure data.  In a second part,  we will address the potential that the same kernel has on \textit{deterministic} data for,  despite having no inherent uncertainty,  improving the model selection.

\section{Gaussian Processes\label{sec:GP}}
Gaussian processes provide a probabilistic approach to regression problems where parameters, linked to a specific model,  are updated as more information is obtained. Formally, they can be described as ``\textit{[...] a collection of random variables, any finite number of which have a joint Gaussian distribution}''\autocite{Book:Rasmussen2005}. 
\par
A Gaussian process is, in essence, a generalisation of a multivariate Gaussian probability distribution for an infinite amount of variables which give us a continuous mean and variance over a feature space (or input coordinate space) with an arbitrary number of features (or dimensions).  With the specification of a kernel,  which is described later in this section,  a GP implies a distribution over functions. 
\par
Given $N$ training data points with inputs $\mathbf{X}_t$,  which  define a $N \times N_{\mathrm{features}}$ matrix,  and responses $\mathbf{y}_t$,  which define a vector of length $N$,  one can then write the joint Gaussian distribution of the latter and a set of query points,  denoted $\mathbf{X}_{*}$,  as
\begin{equation}
\begin{bmatrix}
\mathbf{y}_t \\
\mathbf{y}_{*} 
\end{bmatrix} \sim \mathcal{N}\Bigg( \mathbf{0},  
\begin{bmatrix}
\mathrm{K}(\mathbf{X}_{t},  \mathbf{X}_{t}) & \mathrm{K}(\mathbf{X}_{t},  \mathbf{X}_{*})\\
\mathrm{K}(\mathbf{X}_{*},  \mathbf{X}_{t}) & \mathrm{K}(\mathbf{X}_{*},  \mathbf{X}_{*})
\end{bmatrix} \Bigg)
\end{equation}
where K is the kernel function which,  for produces a matrix of size $N_1 \times N_2$ for inputs of size $N_1 \times N_{\mathrm{features}}$ and $N_2 \times N_{\mathrm{features}}$.  The mean of the distribution does not need to be zero,  it is however more convenient and general to define it as such for the prior and is compatible with normalised training data.
\par
Conditioning the joint distribution,  results in the so-called \textit{posterior} distribution of the GP\autocite{Book:Rasmussen2005} which allows one to model the function with the mean of the latter.  For the query points,  one obtains the predicted means as
\begin{equation}
y (\mathbf{X}_*) = \mathbf{K}_{*t} \ \mathbf{K}^{-1}_{tt} \ \mathbf{y}
\label{eq:gp-maineq}
\end{equation}
where $\mathbf{K}_{*t}$ and $\mathbf{K}_{tt}$ are covariances matrices whose elements are given by the covariance of the point of each dataset (${}_*$ for the query points and ${}_t$ for the training data).  The function $y$,  also called the latent function of the GP,  is the model one uses.  Moreover,  since GPs are probabilistic regression models,  one can obtain a standard deviation of the latent function defined,  at a query point,  as
\begin{equation}
\Delta (\mathbf{X}_*) = \mathbf{K}_{**} \ - \mathbf{K}_{*t} \ \mathbf{K}_{tt}^{-1} \ \mathbf{K}_{t*}
\label{eq:gp-maineq2}
\end{equation}
which is associated with a certainty or confidence on the model prediction.

\section{Kernels}
The kernel which defines the correlation between point in the feature space and the model selection is usually given by two ingredients\footnote{One can use convoluted and product kernels to make much more complex models. We are however here limiting to a single specification of a kernel.}: the main kernel and the noise kernel.  The former will here be given as a scaled (in the sense that an parametrised amplitude is added) Mat\'{e}rn class kernel defined as 
\begin{equation}
\mathrm{K}(\mathbf{X},\mathbf{X}') = \sigma^2 \ \frac{2^{1-\nu}}{\Gamma(\nu)}\Bigg(\sqrt{2\nu}\frac{d}{\rho}\Bigg)^\nu K_\nu\Bigg(\sqrt{2\nu}\frac{d}{\rho}\Bigg) 
\label{eq:covariance-Mat\'{e}rn}
\end{equation}
where $\Gamma$ is the gamma function,  $K_\nu$ is the modified Bessel function of the second kind,  $\rho$ is the length scale of the problem and $d$ is the Euclidean distance $|\mathbf{X} - \mathbf{X}'|$. The main parameter of the Mat\'{e}rn kernel is $\nu$ which is not optimised as it defines the smoothness of the resulting function since a Gaussian process with a Mat\'{e}rn kernel of parameter $\nu = n + 0.5$ will be differentiable $n$ times. The other parameters of the Mat\'{e}rn kernel are optimised (and should be called a hyperparameter as a consequence) and represents the length scales of the problem and the amplitude, $\sigma^2$.  Mat\'{e}rn-based kernels with $\nu=2.5$ are relevant for us as they ensure that the resulting predicted mean of the distribution remains physical in the sense that both the atomic forces (first derivative) and atomic Hessians (second derivative) are smooth.  Moreover it can be noted that for $\nu=\infty$,  an infinitely differentiable function,  on obtain a radial basis function kernel:
\begin{equation}
\mathrm{K}(\mathbf{X},\mathbf{X}') = \sigma^2 \  \mathrm{exp} \Bigg( -\frac{|\mathbf{X} - \mathbf{X}'|}{\rho^2} \Bigg) 
\label{eq:covariance-rbf}
\end{equation}
giving the smoothest latent function within a Mat\'{e}rn descriptor.
\par
In order to allow flexibility of the GP modelling,  a noise component is usually added to the kernel.  This can take many forms but is often a simple White Kernel (WK) which adds a scaled identity matrix to the covariance matrix of the training set to itself as
\begin{equation}
\mathbf{K} \to \mathbf{K} + \lambda^2 \mathbf{I}
\end{equation}
where $\mathbf{K}$ is a matrix of the covariance of $N$ inputs to themselves (K$(\mathbf{X}, \mathbf{X})$ where $\mathbf{X}$ is a $N \times N_{\mathrm{features}}$ matrix),  and where $\lambda$ is a ``noise scaling'' hyperparameter that is optimised alongside other hyperparameter.  The greater the value of $\lambda$,  the noisier the data is considered and the more the latent function is allowed to move away from the training data.  Putting this kernel in equation \ref{eq:gp-maineq} implies that the standard deviation at the training data itself is homogenous on the feature space (homoscedasticity).  As previously stated,  it is not always a fair assumption that energies certainty are homoscedastic as they might come from stochastic methods.  As an alternative to WK,  one can use a Weighted White Kernel (WWK) which adds a scaled non-identity diagonal matrix to the covariance matrix
\begin{equation}
\mathbf{K}(\mathbf{X},  \mathbf{X}) \to \mathbf{K}(\mathbf{X},   \mathbf{X}) + \lambda^2  \begin{bmatrix}
    w_{1} \\
    \vdots\\
    w_{n}
  \end{bmatrix}
\mathbf{I}
\label{eq:wwk-matrix}
\end{equation}
where $w_i$ are weighted noises associated with the training data. The resulting standard deviation of the Gaussian process is not homogenous,  corresponding to the observed data
\par
We optimise our GPs using the Bayesian approach by maximising the log-marginal likelihood\autocite{Book:Rasmussen2005} .


\section{Learning with stochastic electronic structure data\label{sec:kernelsto}}
Post Hartree--Fock methods allow to better treat the correlation of systems with multiple atoms.  We will not give an extensive overview of current methods (one can find such overviews in the literature\autocite{ESOV} ) but we can separate them as \textit{deterministic} electronic structure methods,  such as the Møller--Plesset perturbation theory\autocite{Moller1934} ,  configuration interaction methods\autocite{Sherrill1999} and all the flavours of CC\autocite{Kummel2003},  and,  on the other hand,  \textit{stochastic} electronic structure methods such as Quantum Monte Carlo\autocite{QMC} .
\par
The former two methods scale quite unfavourably with the size of the system and numerically exact solutions become quickly impossible to obtain.  Stochastic electronic structure methods\autocite{Morales2021} aim to improve the size of the space that can be accessed in a given calculation and bypass the need to calculate every configuration.  Stochastic methods have been shown to be able to access larger systems as well as accelerate algorithms implemented for deterministic methods\autocite{Morales2021}. 

Since electronic structure stochastic methods naturally present themselves with an uncertainty on the final energy,  it is natural to define the weights,  $\mathbf{w}$,  of equation \ref{eq:wwk-matrix} related to the associated error on the energy,  $\boldsymbol{\epsilon}$. Taking the toy system from figure \ref{fig:problem},  we scale the randomly generated noises by $_{\mathrm{WK}}\lambda^2_{\mathrm{opt}}$ and optimise a new Gaussian process with a WWK noise.  
\begin{figure}[H]
	\centering
	\includegraphics[width=0.35\textwidth]{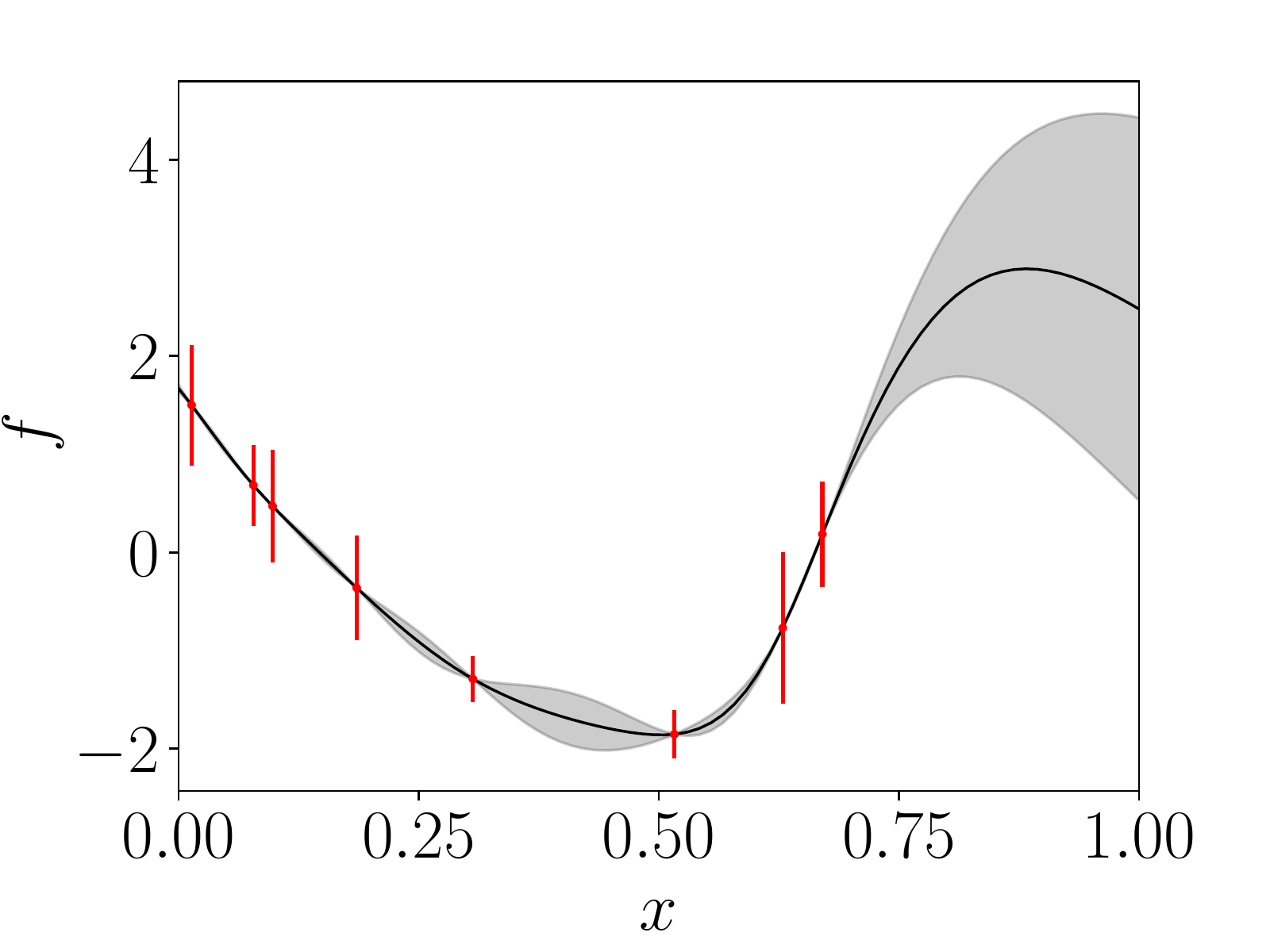}
	\caption[Heteroscedastic kernel optimised for a toy example.]{Latent function (black line) and $\Delta$\textsubscript{95\%} confidence interval (shaded area) of a Gaussian process trained with a WWK on the noisy toy dataset.  Despite the weights being large,  as shown by the red error bars,  the noise hyperparameter $\lambda$ is optimised to very small value $<10^{-6}$ and no changes are seen on the model other than small numerical changes.}
	\label{fig:wwk_optimised}
\end{figure}
Since the uncertainty of a WWK Gaussian process at the training data is scaled by the kernel hyperparameter $\lambda^2$,  resulting GP models are often not improved as shown in figure \ref{fig:wwk_optimised}.  If the noise hyperparameter becomes large,  models become very noisy and do not represent the target function and,  on the other hand,  if $\lambda$ is very small,  the WWK model is essentially the same as a low noise WK model.  The ratios of the confidence of the GP at various training points will respect the ratio of their true uncertainty but will not necessarily reflect its value.  Figure \ref{fig:wwk_optimised} shows the second case and,  at the training data,  the confidence is given by $\lambda^2 w_i$ which disagrees with the true uncertainty,  $w_i$,  of the stochastic data. 
\par
One solution is to assume that the optimised noise hyperparameter of a GP with a WK noise,  $_{\mathrm{WK}}\lambda^2_{\mathrm{opt}}$,  will be similar to the hyperparameter of a GP with a WWK noise.  One can then predict the confidence,  $\lambda^2 w_i$,  of the latter and define new scaled weights as $w'_i = {}_{\mathrm{WK}}\lambda^2_{\mathrm{opt}} w_i$.  The confidence of the WWK model is now,  still assuming the $\lambda$ hyperparameters are optimised to similar values,  given by
\begin{equation}
 _{\mathrm{WWK}}\lambda^2_{\mathrm{opt}} w'_i =   \frac{_{\mathrm{WK}}\lambda^2_{\mathrm{opt}}}{_{\mathrm{WWK}}\lambda^2_{\mathrm{opt}}} w_i \simeq w_i
\label{eq:wk-wwk-agreement}
\end{equation}
which agrees with the original uncertainty on the data\footnote{If the WK and WWK $\lambda$ hyperparameters have a ratio that is far from unity,  one will not see an agreement with the original data.  One can either reassess the weights and increase or decrease them until equation \ref{eq:wk-wwk-agreement} produces satisfactory agreement or one can set a non-optimisable WWK noise with ${}_{\mathrm{WK}}\lambda^2={}_{\mathrm{WWK}}\lambda^2$,  making the uncertainty equal.  The latter is not preferable and it was not used in these results.}.
\par
\begin{figure}[H]
	\centering
	\subfloat[WK]{\includegraphics[width=0.2\textwidth,trim=0cm 0 1cm 0, clip]{graphics/gp_wn_normal}}\hfill
	\subfloat[WWK]{\includegraphics[width=0.2\textwidth,trim=0cm 0 1cm 0, clip]{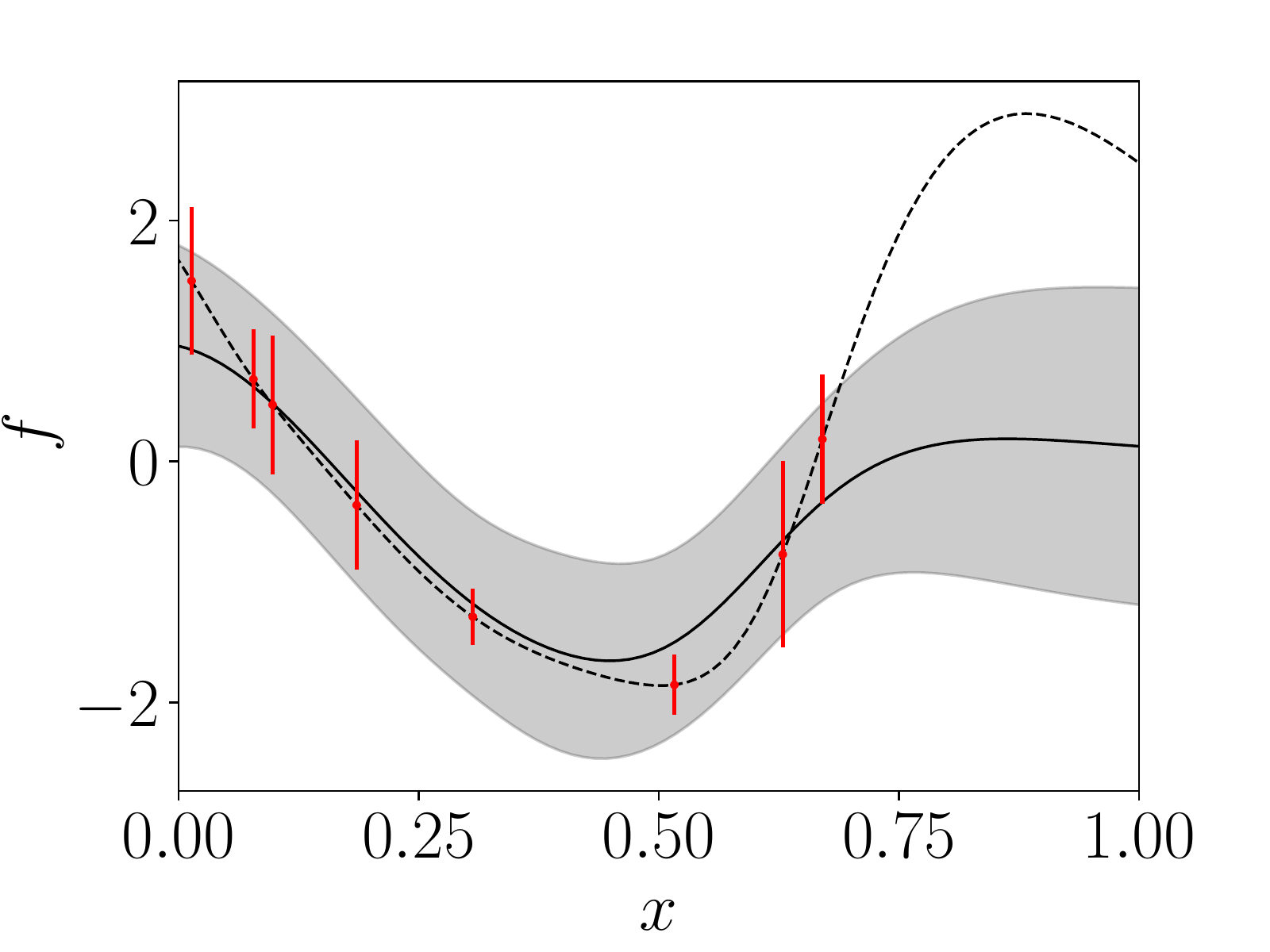}}\hfill
	\caption[Smoothing of the Gaussian process latent function resulting from the use of a weighted noise kernel.]{Latent function (black line) and $\Delta$\textsubscript{95\%} confidence interval (shaded area) of the Gaussian process with a WK and a WWK,  respectively,  with weights,  given by $\mathbf{w}/_{\mathrm{WK}}\lambda^2_{\mathrm{opt}}$,  are randomly generated noises (shown by the red error bars) and $_{\mathrm{WK}}\lambda^2_{\mathrm{opt}}$ is the optimised hyperparameter for the WK,  resulting in the model in the first panel.  The latent function of the WK Gaussian process is plotted in the left panel as a dashed line for comparison purposes.  The smoothing of the latent function for the WWK is very strong but expected as for visual purposes the noises were very large.  As this is a toy system there is no metric to say which of the two curves in panel (b) is closer to the target function and all we are trying to show,  is the effect of noise on the training data when a WWK is used.}
	\label{fig:fake-sto}
\end{figure}

The new latent function,  and standard deviation,  is shown in figure \ref{fig:fake-sto},  where two distinctive effects are seen:  the uncertainty on the predictions is not vanishingly small at the training data but remains consistent with the original uncertainty.  Moreover,  the length scale of the Mat\'{e}rn ($\nu=2.5$) becomes larger,  as expected for noisier data,  giving a smoothing effect on the latent function.  Both of these can be clearly seen in figure \ref{fig:fake-sto} but quite exaggerated due to the large exaggerated noises that were used on the toy system.
\par
\begin{figure}[H]
	\centering
	\subfloat[WK]{\includegraphics[width=0.2\textwidth,trim=0cm 0 1cm 0, clip]{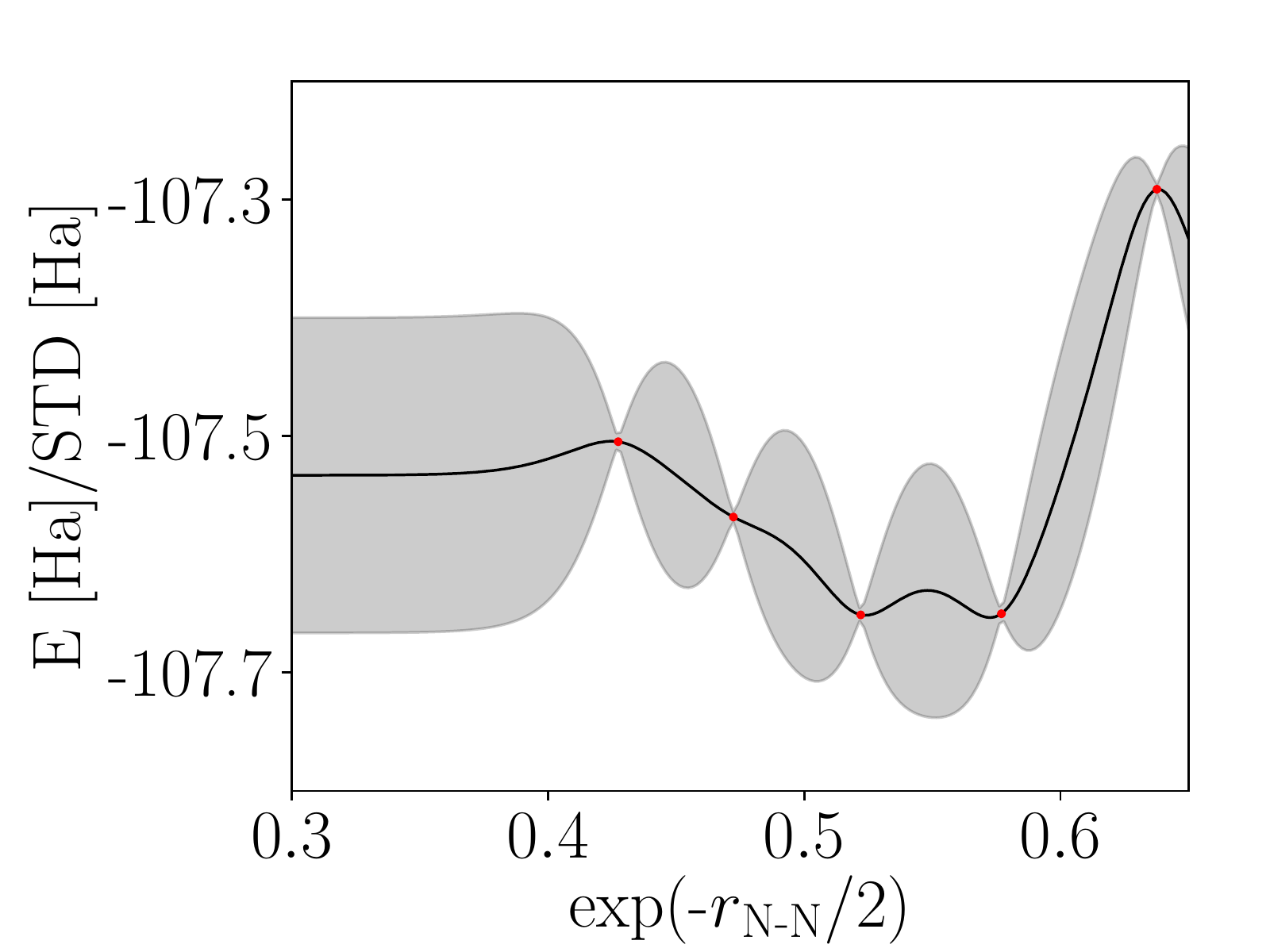}}\hfill
	\subfloat[WWK]{\includegraphics[width=0.2\textwidth,trim=0cm 0 1cm 0, clip]{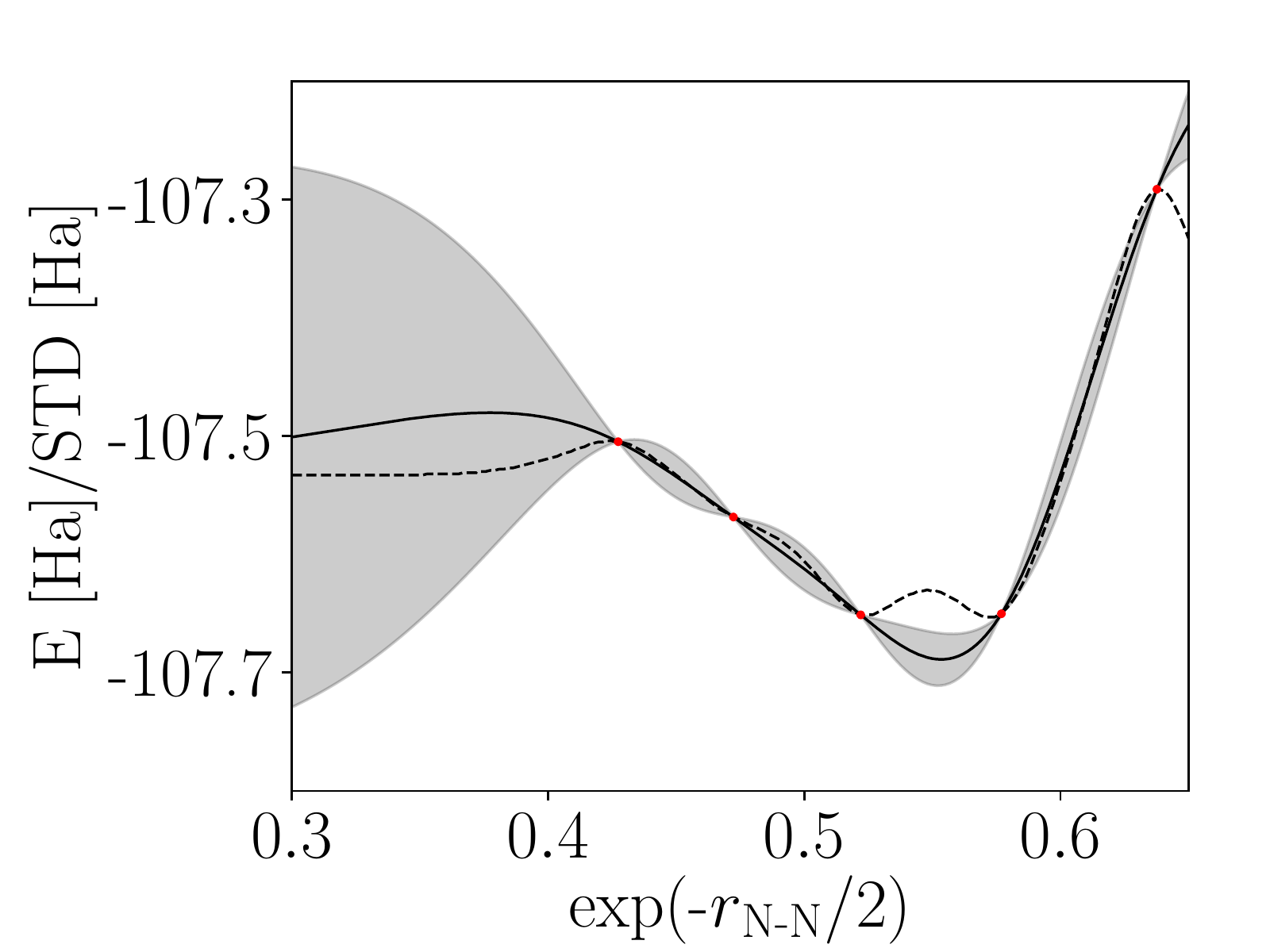}}\hfill
	\caption[Weighted noise kernel Gaussian process results on an electronic structure dataset of N$_2$ energies.]{Latent function (black line) and $\Delta$\textsubscript{95\%} confidence interval (shaded area) of the Gaussian process with a WK and a WWK,  respectively,  for the N$_2$ training set.  The latent function of the WK Gaussian process is plotted in the right panel as a dashed line for comparison purposes. }
	\label{fig:n2-sto}
\end{figure}

In order to asses the WWK performance over realistic electronic structure data,  a training set of coupled-cluster Monte Carlo (CCMC) energies\autocite{Thom2010, Spencer2016, Scott2017, Scott2020} ,  CCMCSD/STO-3G of N$_2$ calculated with HANDE\autocite{HANDE} ,  is modelled with a simple WK Gaussian process.  A typical regime of over fitting can be observed (see panel (a) of figure \ref{fig:n2-sto}) with a very short length scale and a latent function that does not correspond to the true \textit{ab initio} function. 
\par
On the other hand,  one can use the error bars of the CCMCSD calculations to assign the weights,  $\mathbf{w}$,  scale them with the $\lambda$ of the WK (of the model shown in figure \ref{fig:n2-sto}) and train a GP with a WWK kernel.  The latter produces a net improvement in the latent function,  as shown in panel (b) of figure \ref{fig:n2-sto}. 
\par
The tightening of the standard deviation around the training values is still visually observed here.  However,  rather than converging to the value of $_{\mathrm{WK}}\lambda^2_{\mathrm{opt}}$,  they converge to the value of $(_{\mathrm{WK}}\lambda^2_{\mathrm{opt}}/_{\mathrm{WWK}}\lambda^2_{\mathrm{opt}}) w_i \sim \epsilon_i$ which is more realistic (in this instance the two optimised $\lambda$ hyperparameters are almost identical).  The smoothing effect is much more interesting as it allows the latent function of the WWK Gaussian process to ``escape'' the over fitting regime and produce a much more realistic curve. 

\par
\begin{figure}[H]
	\centering
	\subfloat[WK]{\includegraphics[width=0.2\textwidth,trim=0cm 0 1cm 0, clip]{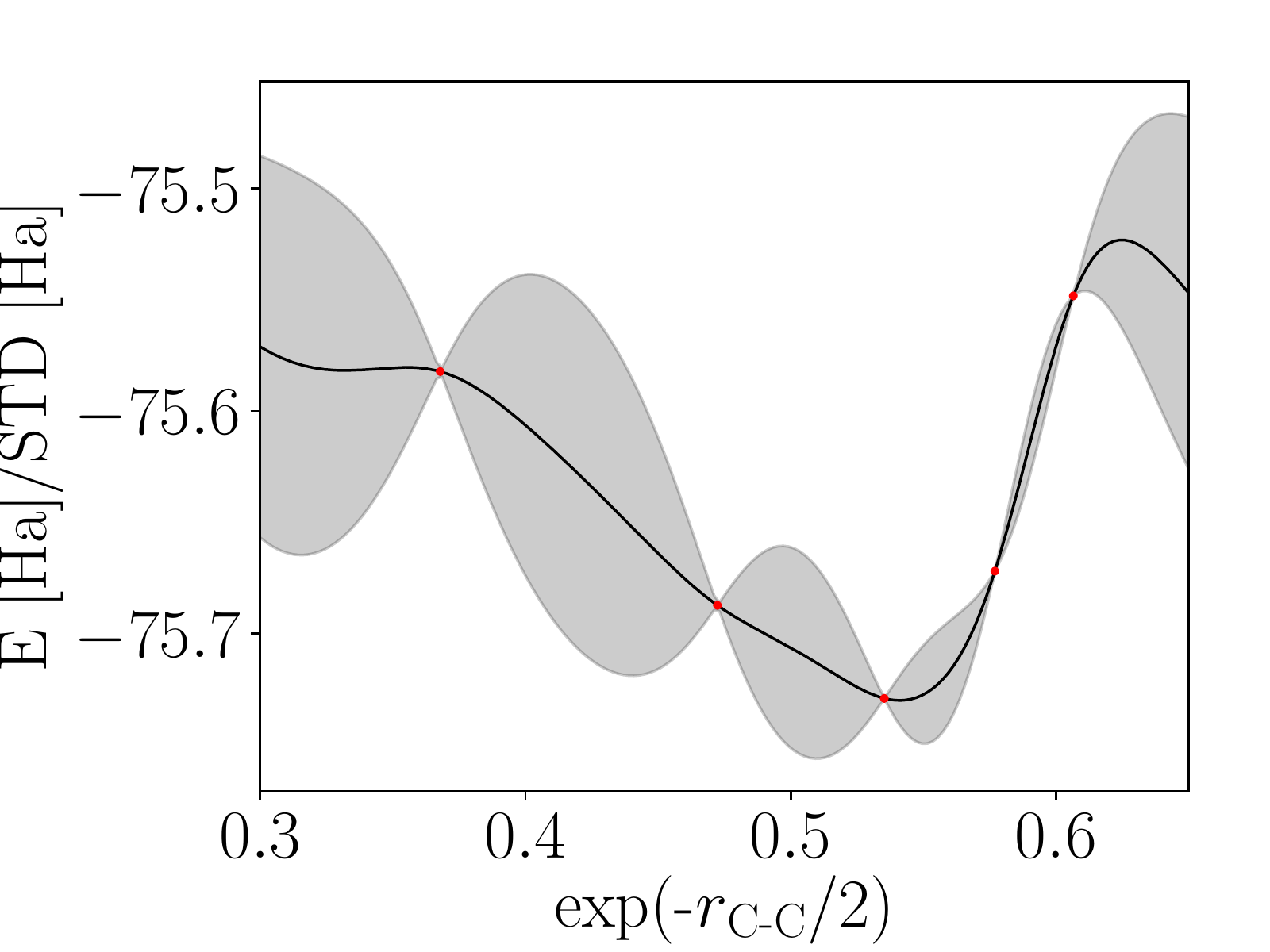}}\hfill
	\subfloat[WWK]{\includegraphics[width=0.2\textwidth,trim=0cm 0 1cm 0, clip]{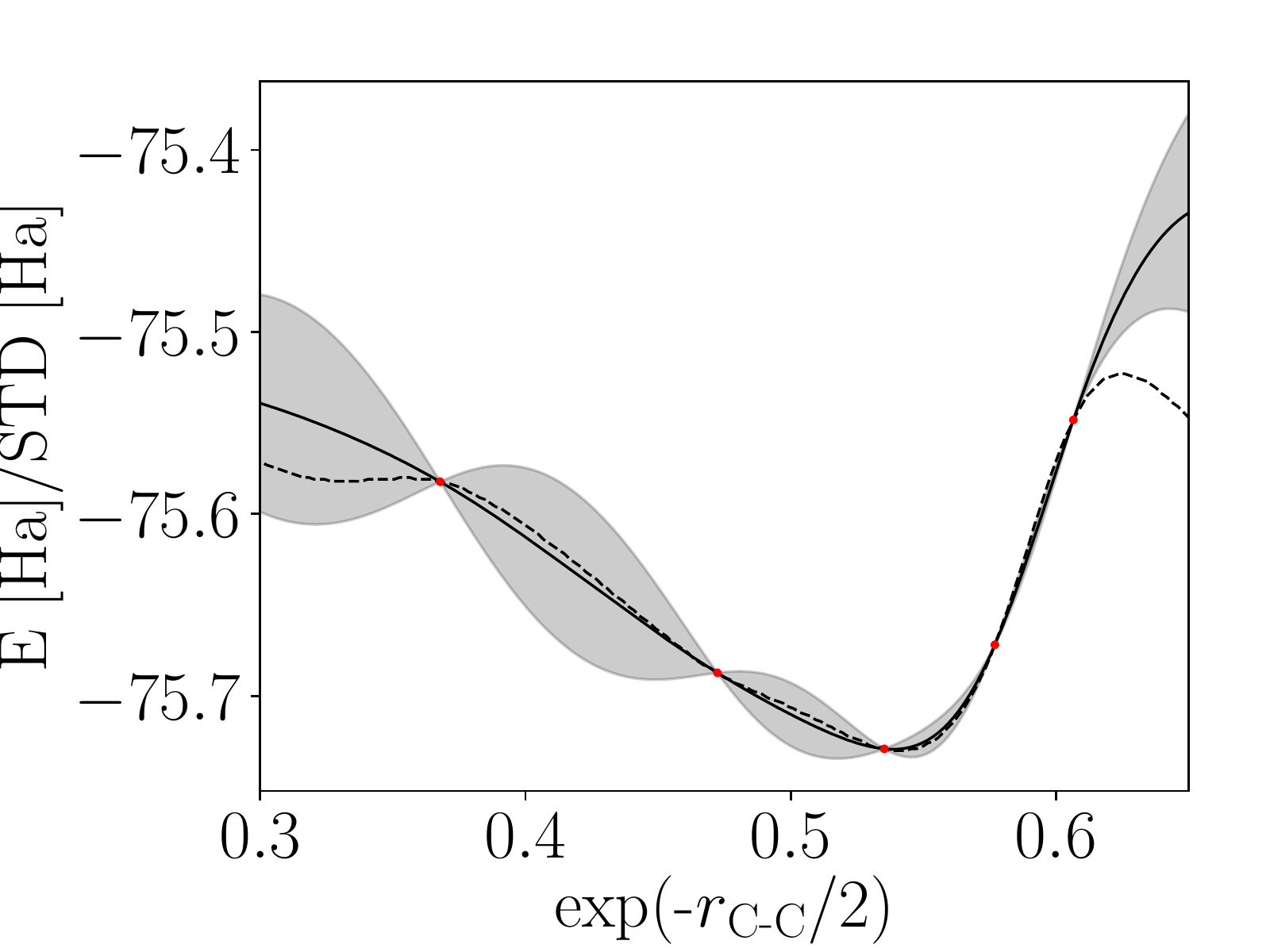}}\hfill
	\caption[Weighted noise kernel Gaussian process results on an electronic structure dataset of C$_2$ energies.]{Latent function and standard deviation of the Gaussian process with a WK and a WWK,  respectively,  for the C$_2$ training set.  The latent function of the WK Gaussian process is plotted in the right panel as a dashed line for comparison purposes. }
	\label{fig:c2-sto}
\end{figure}

As a second important example,  we use a dataset derived from the FCIQMC/6-31G$^{*}$ C$_2$ energies given by  Booth \textit{et al.}\autocite{Booth2011},  where a simple Gaussian process with a WK does not over fit (see figure \ref{fig:c2-sto}),  in order to assess if the smoothing is not too excessive in this regime.  However,  the $_{\mathrm{WWK}}\lambda^2_{\mathrm{opt}}$ happens to be much larger than the $_{\mathrm{WK}}\lambda^2_{\mathrm{opt}}$ for this system,  yielding larger noises and a very noisy final model.  To address this issue a further rescaling is done to obtain smaller weights and a $_{\mathrm{WWK}}\lambda^2_{\mathrm{opt}} \sim {}_{\mathrm{WK}}\lambda^2_{\mathrm{opt}}$ is found,  yielding the following latent functions.

One can see that again model smoothing happens but not on the same scale as the N$_2$ dataset,  as the original latent function was much smoother.  Although some user input is needed to find the correct scaling (a process which can be automated),  using WWK in Gaussian processes allows one to improve modelling of potential energy surfaces and,  for stochastic data,  allows the latent function of the model to reflect the true uncertainty of the data.

\section{Learning with deterministic electronic structure data\label{sec:kerneldet}}

Despite deterministic data not having an uncertainty that contradicts the standard deviation of a Gaussian process,  the smoothing effect of WWK offers a potential useful tool for improving modelling,  \textit{i.e.} improving Gaussian processes in the over fitting regime.  However,  since there is no intrinsic noise that can be associated to each data to create the weights needed in equation \ref{eq:wwk-matrix},  one has to use \textit{constructed} noises.  These offer ways to bias the learning process towards ``importance'' and improve the latent function.  First,  we recall the model selection process of Gaussian processes,  which is done by maximising the log-marginal likelihood\autocite{Book:Rasmussen2005} given by
\vfill
\begin{equation}
\begin{aligned}
\mathrm{LML} = -&\frac{1}{2} \mathbf{y}^{\mathrm{T}} \mathbf{K}^{-1} \mathbf{y} - \frac{1}{2} \mathrm{log} |\mathbf{K}| -\\[0.3cm]
 &\frac{n}{2} \mathrm{log} (2\pi)
\end{aligned}
\end{equation}
\vfill
Over fitting regime is: when the \textit{data} term dominates the log-marginal likelihood and pushes models with high complexity to higher values.  It is rather hard to slightly bias model selection to a lower complexity model for a given dataset and,  what is usually seen,  is a set of competing minima on the LML surface,  which are either \textit{data} term-dominated or \textit{complexity} term-dominated.  The former is in the over fitting regime while the latter have a large $\lambda$ and are not models that describe the data in a \textit{quantitative} manner.  This remains true for a WWK Gaussian process.
\par
If one has a very large weight in the noise model for a given data point,  the rough effect is that the \textit{data} term of the LML has terms that scale with $\mathrm{f}_i^2 w_i^{-1}$ since the inverse of the covariance matrix is taken\footnote{Obviously, the inverse is not simply given by the inverse of the diagonal weight matrix but for the sake of simplicity we use that result.}.  As a consequence,  the larger the noise is the less the data of point $i$ is relevant and the more the \textit{complexity} term dominates the LML value,  smoothing the models.
\par
Using the same dataset of N$_2$ CCMCSD/STO-3G energies (we now pretend these are deterministic results with only a very small numerical noise on them) which over fits the training data with a WK Gaussian process,  as shown in in figure \ref{fig:n2-sto},  we construct a bias model with exponential weights to create noise on high energy data and force a better fit on the bottom of the PEC.  We work here with multiple constructed noises given by $\epsilon_i = \mathrm{exp}(\beta \mathrm{E}^{*}_i)$ where $\mathrm{E}^{*}_i= \mathrm{E}_i - \mathrm{min}(\mathbf{E})$.  
\begin{figure}[H]
	\centering
	\includegraphics[width=0.4\textwidth]{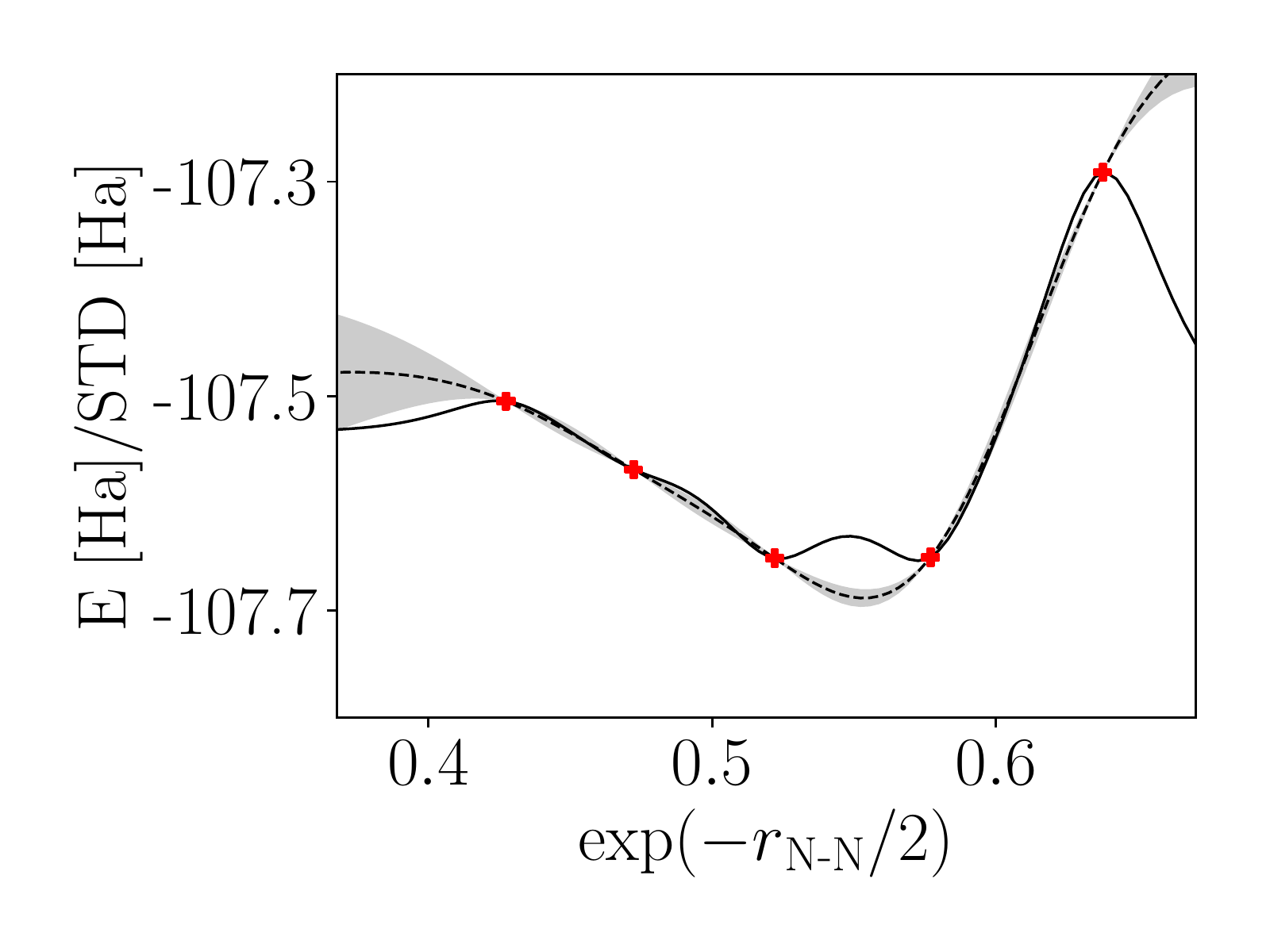}
	\caption[Gaussian process models with different scaling of the $\beta$ parameter.]{Latent function (black line) and $\Delta$\textsubscript{95\%} confidence interval (shaded area) of a Gaussian process trained with a biased WWK with exponential weights and a parameter set to $\beta=0$ on the N$_2$ CCMCSD/STO-3G dataset.  The true function is shown by the dashed line.}
	\label{fig:n2-model0}
\end{figure}
At $\beta=0$,  the WWK has all weights equaling 1 and is thus equivalent to the standard WK.  As expected,  the latent function of the GP is the same as the one shown in figure \ref{fig:n2-sto}(b) and over fitting of the training data is present.  Once the $\beta$ parameter is increased,  the models are no longer equivalent to the GP with a WK noise.

\begin{figure}[H]
	\centering
	\subfloat[$\beta=1/3$]{\includegraphics[width=0.22\textwidth,trim=0cm 0 0.5cm 0, clip]{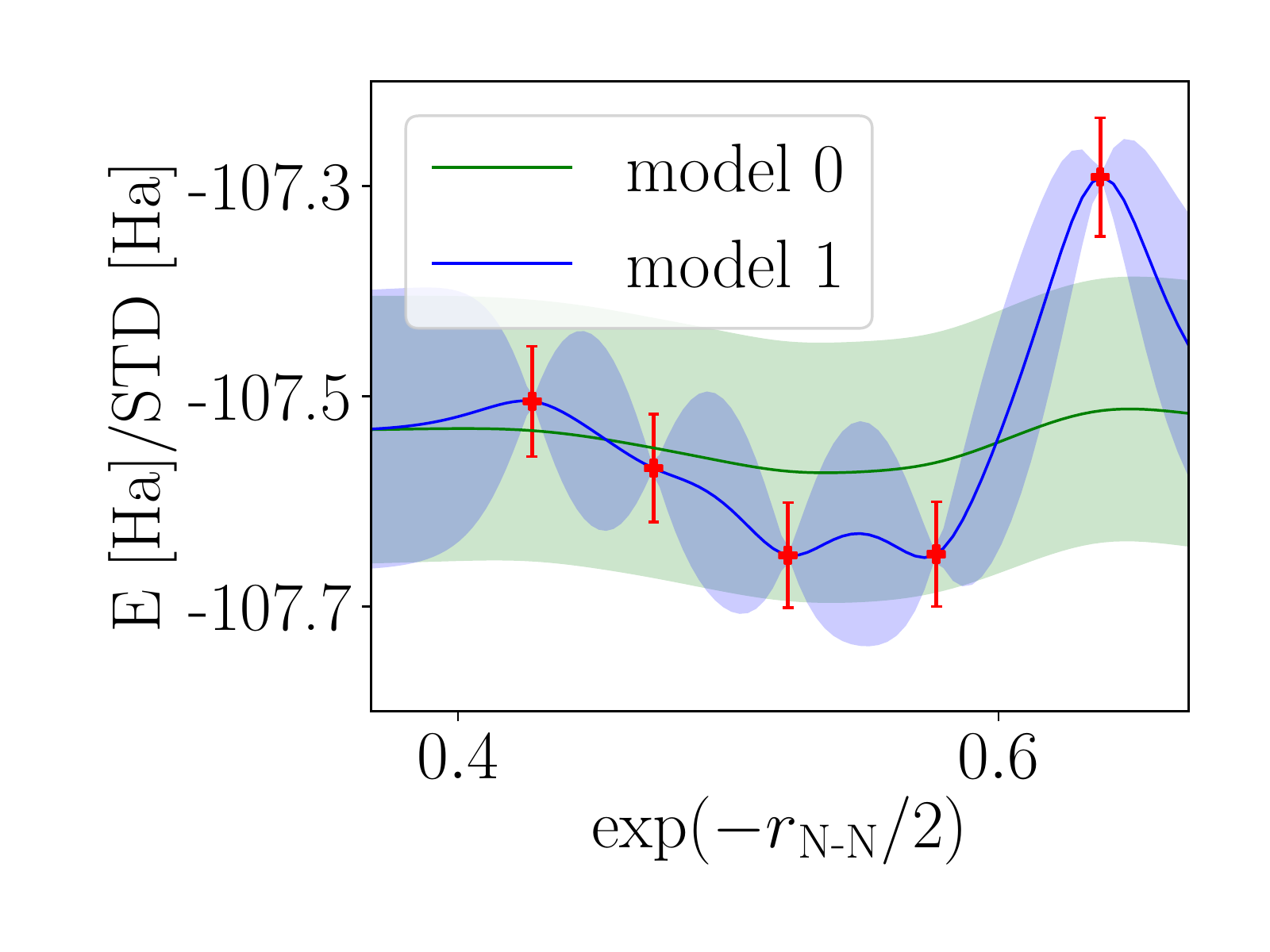}}\hfill
	\subfloat[$\beta=1/2$]{\includegraphics[width=0.22\textwidth,trim=0cm 0 0.5cm 0, clip]{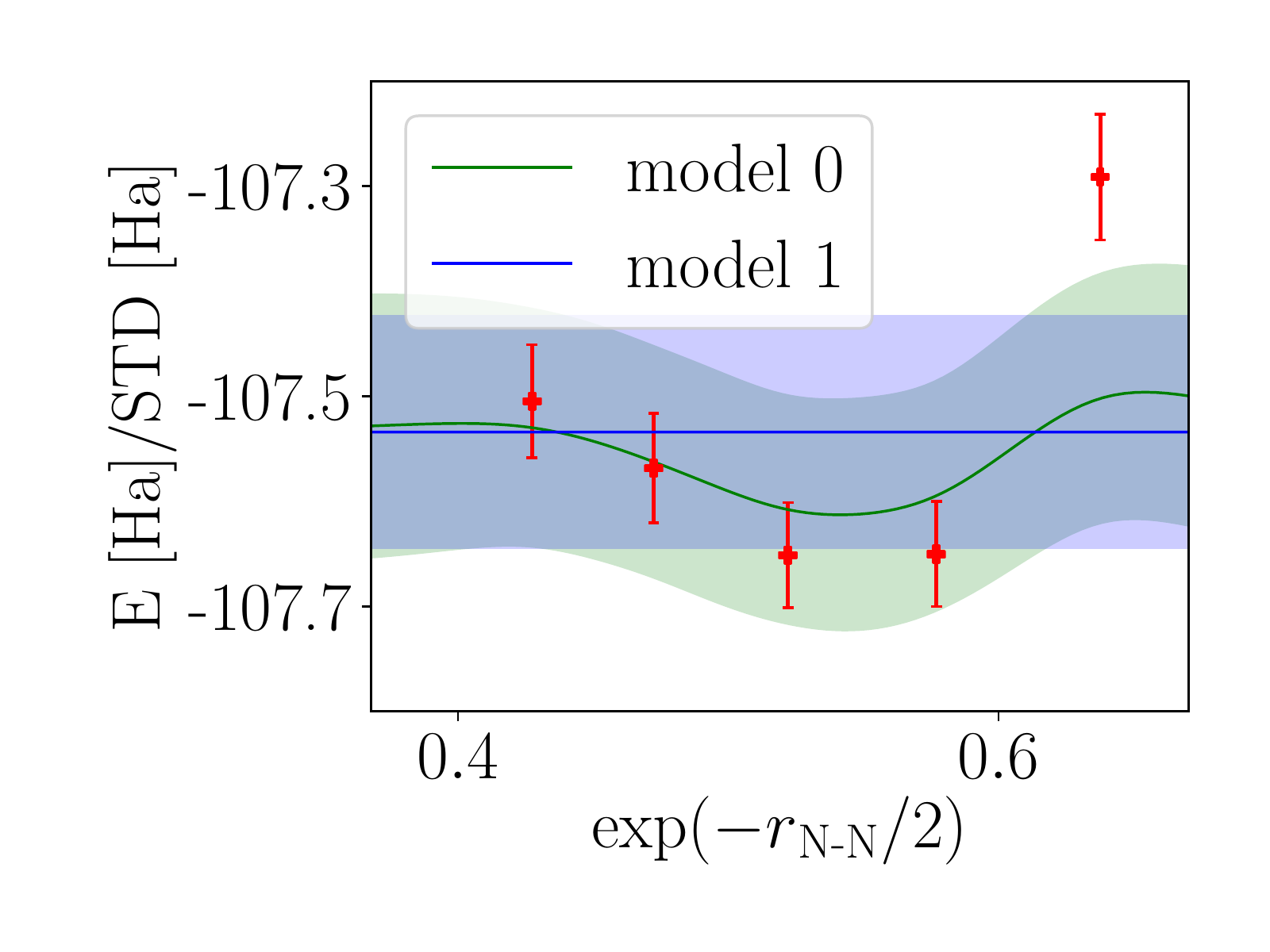}}\hfill
	\subfloat[$\beta=5$]{\includegraphics[width=0.22\textwidth,trim=0cm 0 0.5cm 0, clip]{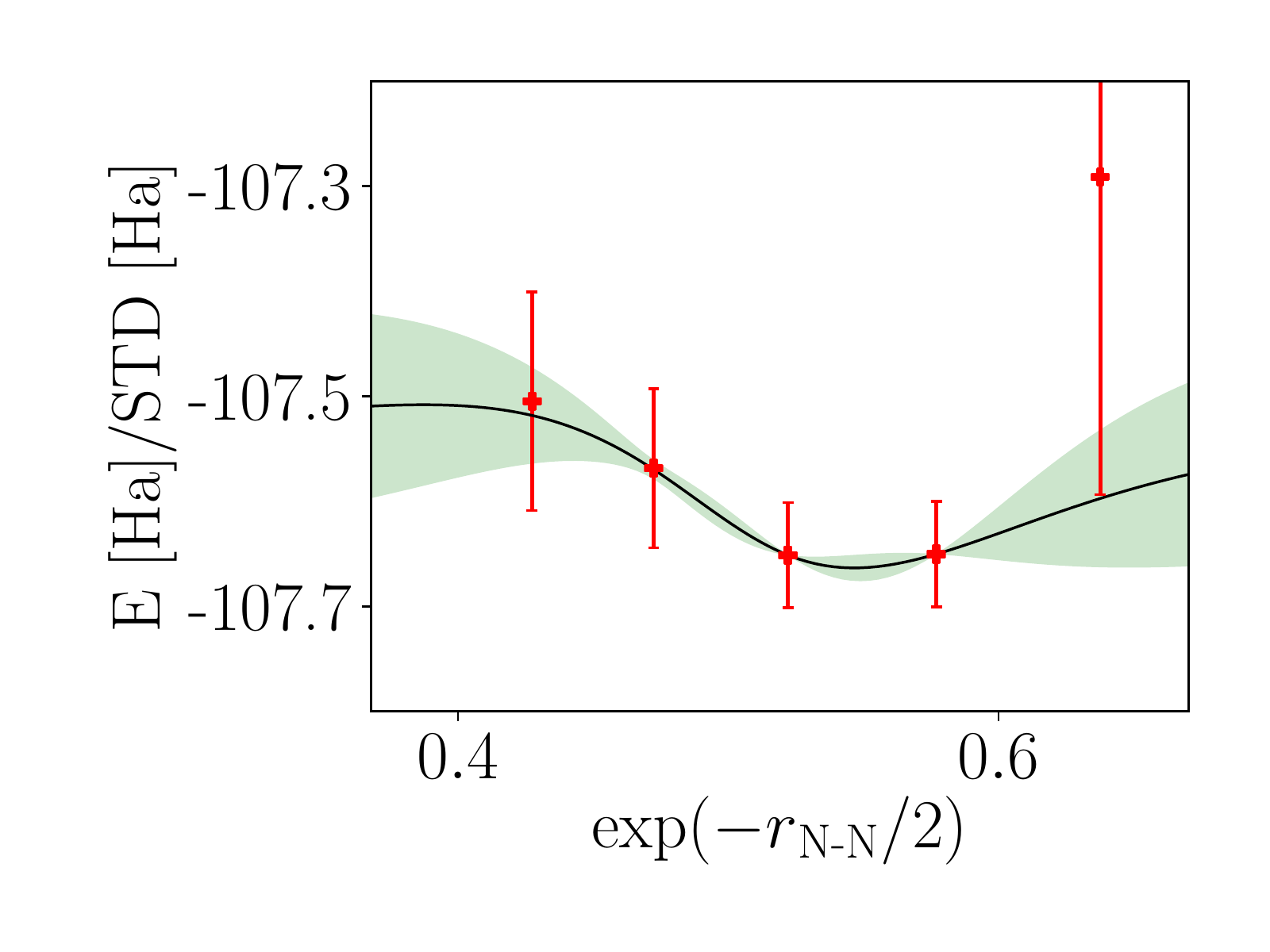}}\hfill
	\caption[Gaussian process models with different scaling of the $\beta$ parameter.]{Progression of WWK Gaussian process models,  each corresponding to a local minimum on the LML surface and numbered from the global minimum which has its associated model 0 (the numbering follows the LML of each GP and is not following a minimum through the different noise models).  These results show GP trained on biased WWK noises for parameters  set to $\beta>0$ (the associated errors bars are shown for $\lambda^2=0.05$) and trained on the N$_2$ CCMCSD/STO-3G dataset.}
	\label{fig:n2-all-models}
\end{figure}
Three main regimes of WWK scaling are outlined in figure \ref{fig:n2-all-models}: for small $\beta$ values,  a new model,  corresponding to the \textit{complexity} term-dominated model,  appears while the original WK model is still present.  At intermediate $\beta$ values,  all models are \textit{complexity} dominated and very noisy and,  surprisingly,  the original WK model is no longer seen.  Eventually,  as $\beta$ is even larger,  a single model with a better fit of the dataset (except the high energy point which has a large noise associated to it) is seen.  The reason for this observation is that a large $\beta$ makes the ratio of weights,  $w_i$,  large allowing small $\lambda$ values to still model a large noise on high energy data,  given by $\lambda w_i$.
\par
\begin{figure}[H]
	\centering
	\subfloat[WK optimised $\boldsymbol{ \theta}$]{\includegraphics[width=0.2\textwidth,trim=0cm 0 0.5cm 0, clip]{graphics/N2_models_0}}\hfill
	\subfloat[WWK optimised $\boldsymbol{ \theta}$]{\includegraphics[width=0.2\textwidth,trim=0cm 0 0.5cm 0, clip]{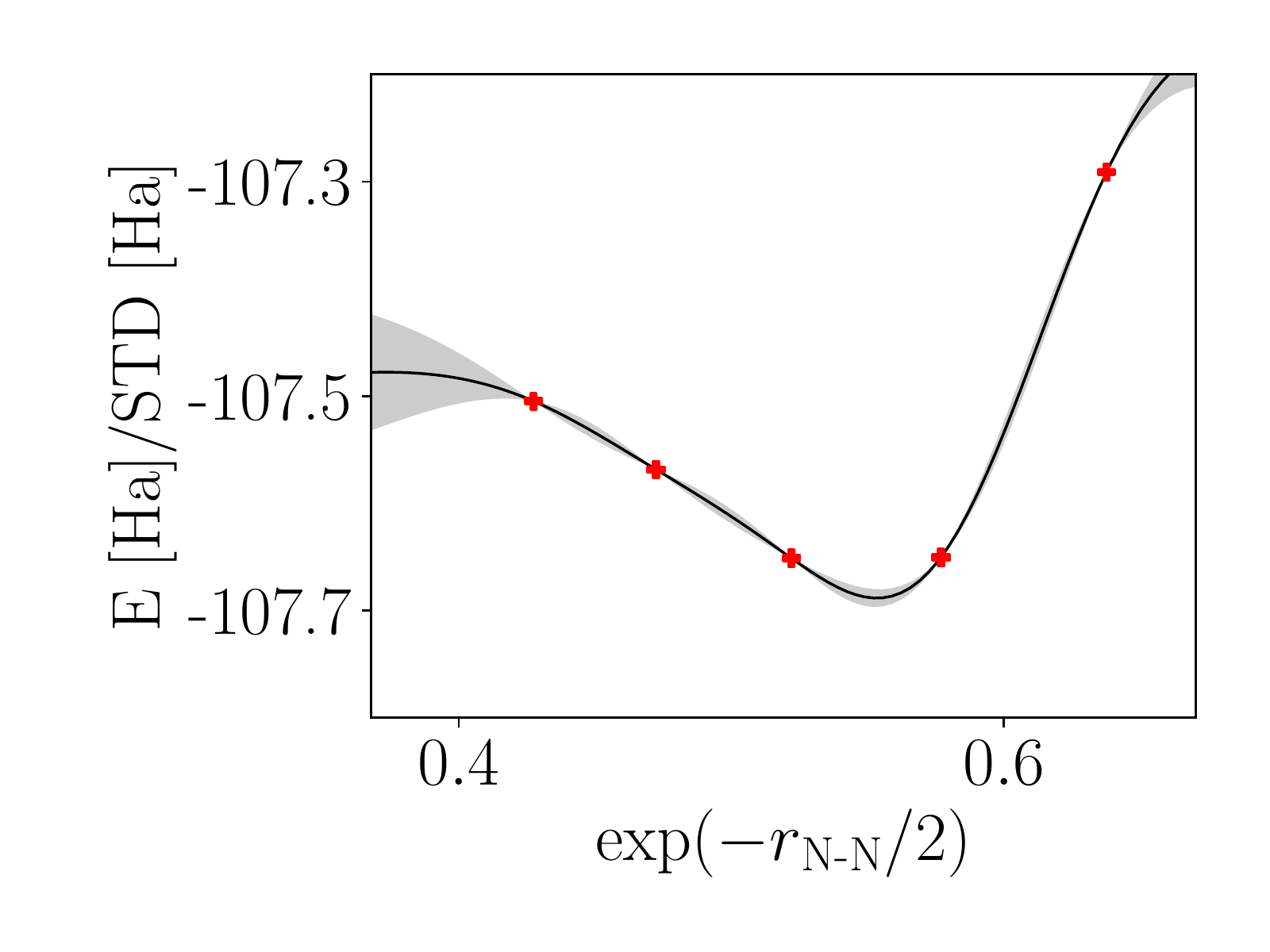}}\hfill
	\caption[Gaussian process with combined hyperparameters from different noise kernels.]{Resulting latent function (black line) and $\Delta$\textsubscript{95\%} confidence interval (shaded area) of WK Gaussian processes with original hyperparameters and WWK hyperparameters.  One can see the strong smoothing effect of the WWK hyperparameters that removes over fitting and produces a much better surface.}
	\label{fig:n2-mixed-learning}
\end{figure}

Gaussian process with a WWK and large weights are not very useful for \textit{quantitative} modelling as the latent function in figure \ref{fig:n2-all-models}(c) clearly shows.  However,  what these models are good at is to optimise to length scales that effectively ``see'' less data and tend to lengthen,  since the \textit{data} term becomes less and less important in the LML.  This is a rather interesting effect for GP that over fit the training data since one can push the GP out of that regime.  However,  ignoring some training data also makes the model miss the target function but if one combine the WWK hyperparameters,  $\boldsymbol{ \theta} = (\sigma, \rho, \lambda)$,  which do not over fit,  with a WK GP and does not reoptimise them one can obtain a model that \textit{sees} all the data in the training set but only used a subset of that data to optimise the model.  This leads to a smoothing effect like the one in figure \ref{fig:n2-all-models}(d) without heavy deviation from the training data,  as seen in figure \ref{fig:n2-mixed-learning}.

\section{Conclusion}
When using stochastic electronic structure methods to create training data,  including the true heteroscedastic noise in the GP learning scheme is not only useful to force the confidence of the model to respect the true error of the electronic structure data but is also useful to improve the model by regularisation of the GP latent functions.  Moreover,  since over fitting is a common problem to regressive learning,  one can take advantage of the regularisation effect of the WWK to create smoother latent functions of deterministic electronic structure data.
\par
This second point is explored by introducing fictitious noises with a ``chemical'' reasoning behind it: in this instance allowing the GP to be less accurate on the high energy data,  biases the learning toward simpler models.  The heteroscedastic noise is equivalent to setting ``importance weights'' on the data: large noises implying a small importance in the learning process.  This does not need to permanently remove the high energy data from the model,  as one can see in figure \ref{fig:n2-mixed-learning}.  The latter can be reintroduce in a new GP with homoscedastic noise and pre-optimised hyperparameters to yield a latent function.  The high energy data will now impact the GP latent function but the covariance spans with the length scale optimised on the low energy data only.

\section*{Acknowledgments}
I would like to thank the Royal Society for funding as well as the Wales group of the University of Cambridge for providing access to the GMIN suite\autocite{GMIN} .

\printbibliography
\end{multicols}
\end{document}